\documentclass[12pt,preprint]{aastex}
\usepackage{emulateapj5}
\usepackage{natbib,psfig}

\slugcomment{To appear in ApJ Letters}

\shorttitle{An FR\,I radio structure associated with the optically-powerful
quasar E\,1821+643}
\shortauthors{Blundell \& Rawlings}

\begin{document}

\title{The optically-powerful quasar E\,1821+643 is associated with a
300-kpc scale FR\,I radio structure}

\author{Katherine M.\ Blundell\altaffilmark{1} and Steve Rawlings\altaffilmark{1}}

\altaffiltext{1}{University of Oxford, Astrophysics, Keble Road, Oxford,
OX1 3RH, U.K.}

\begin{abstract}
We present a deep image of the optically-powerful quasar E\,1821+643
at 18\,cm made with the Very Large Array (VLA).  This image reveals
radio emission, over 280\,$h^{-1}$\,kpc in extent, elongated way
beyond the quasar's host galaxy.  Its radio structure has decreasing
surface brightness with increasing distance from the bright core,
characteristic of FR\,I sources \citep{Fan74}.  Its radio luminosity
at 5\,GHz falls in the classification for `radio-quiet' quasars
\citep[it is only $10^{23.9}\,{\rm W Hz^{-1} sr^{-1}}$ see
e.g.][]{Kel94}.  Its radio luminosity at 151\,MHz (which is
$10^{25.3}\,{\rm W Hz^{-1} sr^{-1}}$) is at the transition luminosity
observed to separate FR\,Is and FR\,IIs.  Hitherto, no
optically-powerful quasar had been found to have a conventional
FR\,I radio structure.  For searches at low-frequency this is
unsurprising given current sensitivity and plausible radio spectral
indices for radio-quiet quasars.  We demonstrate the inevitability of
the extent of any FR\,I radio structures being seriously
under-estimated by existing targetted follow-up observations of other
optically-selected quasars, which are typically short exposures of $z
> 0.3$ objects, and discuss the implications for the purported radio
bimodality in quasars. 

The nature of the inner arcsec-scale jet in E\,1821+643, together with
its large-scale radio structure, suggest that the jet-axis in this
quasar is precessing (cf.\ Galactic jet sources such as SS\,433).  A
possible explanation for this is that its central engine is a binary
whose black holes have yet to coalesce.  The ubiquity of precession in
`radio-quiet' quasars, perhaps as a means of reducing the observable
radio luminosity expected in highly-accreting systems, remains to be
established.
\end{abstract}


\section{Introduction}
\label{sec:intro}

Radio sources from the 3CRR sample \citep{Lai83} whose nuclei are
quasars tend to be associated with classical double [FR\,II,
\citet{Fan74}] structures.  This sample contains no examples of radio
sources clearly having the other characteristic structure, the FR\,I
type (where the brightness decreases with increasing distance from the
core), powered by quasar nuclei.  This, together with a consistent
picture from similar samples, led some to postulate that there is a
deep association between the nature of the central engine and whether
the radio-structure is type FR\,I or FR\,II \citep{Bau95}.

In principle, a survey like 3CRR is hindered from finding FR\,I radio
structures associated with quasars because of two effects.  First, as
pointed out by \citet{Fan74}, FR\,Is have lower radio luminosities
than the classical double FR\,IIs (see Fig.\,\ref{fig:rqq_fr}); this
means they are not found at high redshifts in a bright flux-limited
sample.  Second, the rarity of quasars means that searching the
requisite large volume of space necessarily involves going out to high
redshifts.

An alternative means of finding quasars associated with FR\,I radio
structures is to begin with an optical survey of quasars and then
make follow-up radio observations to establish the nature of their
structures.  In the case of the BQS sample of quasars \citep{Sch83}
the radio structures, when prominent, are only of the classical double
FR\,II type.  Observations to date show the other quasars appear to
have weak, though compact \citep{Blu98}, radio emission associated
with their nuclei.  This indicates a seeming gap in radio luminosity
for quasars of a given optical luminosity, as pointed out by e.g.\
\citet{Mil93}: at radio luminosities corresponding to the FR\,Is from
the 3CRR sample, there are no BQS quasars.

The absence of evidence of quasars with these radio luminosities (the
radio luminosities of FR\,I radio galaxies in 3CRR) was taken by
\citet{Fal95} as evidence of the absence of these objects in the
Universe.  This led them to postulate that torus opening angles in
FR\,Is are too small to observe a quasar nucleus: for most angles to
the line-of-sight, the nucleus would be obscured by the torus, while
for very small angles to the line-of-sight the nuclear emission would
be that of a strongly core-boosted BL Lac.  The wider opening angles
of FR\,II sources permit a significant fraction of these objects to be
observed as quasars.  Thus, the seeming absence of FR\,I quasars was
taken by \citet{Fal95} as evidence that the nature of the torus is
closely linked to the FR\,II/FR\,I transition.

A possible counter to their claim was the discovery by \citet{Lar99}
of an FR\,I radio structure whose nuclear identification, while not a
bona fide quasar, exhibits broad lines.  With a projected linear size
of $\sim 0.5$\,Mpc it seemed most unlikely that this object was being
viewed at a very small, favourable angle to the line-of-sight.
Examples of FR\,I sources having bona fide quasar nuclei would
undermine \citet{Fal95}'s theory.

Deep, low-frequency, radio surveys are a means of sampling the
high-$z$ Universe necessary to find rare quasars.  However, in the
low-radio luminosity regime where FR\,I sources are found, current
surveys remain challenged by sensitivity.  But despite the limited sky
coverage of redshift surveys of deep low-frequency radio sources to
date, a low-radio luminosity source from the 38-MHz selected 8C sample
\citep{Lac99}, which we find to have an FR\,I structure over 280\,kpc
in extent, is associated with a well-studied quasar which is optically
extremely powerful \citep[][unreddened $M_V = -27.5$]{Hut91}.

What are the chances that optically-selected quasars will be found to
have associated FR\,I structures?  Given that optically-powerful
quasars are too rare to populate the nearby Universe and the
consequences of redshift on surface brightness discussed in
\S\ref{sec:redshift}, the short interferometric snapshots which
comprise most observations of these objects to date are inadequate to
reveal any such structures.  Psychology presumably plays its part in
discouraging observers, or at least proposal referees, from making
deep radio observations of seemingly `radio-quiet' quasars.  In
\S\ref{sec:meas} we briefly describe what our deep observations of
E\,1821+643 have revealed.  In \S\ref{sec:results} we discuss the size
and luminosity of its elongated emission and consider the stability in
jet-axis direction that is implied by the large-scale radio emission.
In \S\ref{sec:redshift} we consider the detectability, given current
technology, of the extended emission if the quasar were located at $z
= 1$ instead of its true value $z = 0.297$ \citep{Sch92}.

In \S\ref{sec:sourcephysics} we discuss the challenges this quasar
presents to the implications of the radio bimodality in quasars, the
radio-optical correlation \citep[e.g.][]{Raw91,Wil99} and the assumed
mapping between jet power and observable radio luminosity.

We assume that $h$ is the Hubble constant in units of $50 ~ \rm
km\,s^{-1} Mpc^{-1}$, and that $\Omega_{\mathrm M} = 0.3$ and
$\Omega_{\Lambda} = 0.7$.

\section{Observations}
\label{sec:meas}
\begin{figure}
\plotone{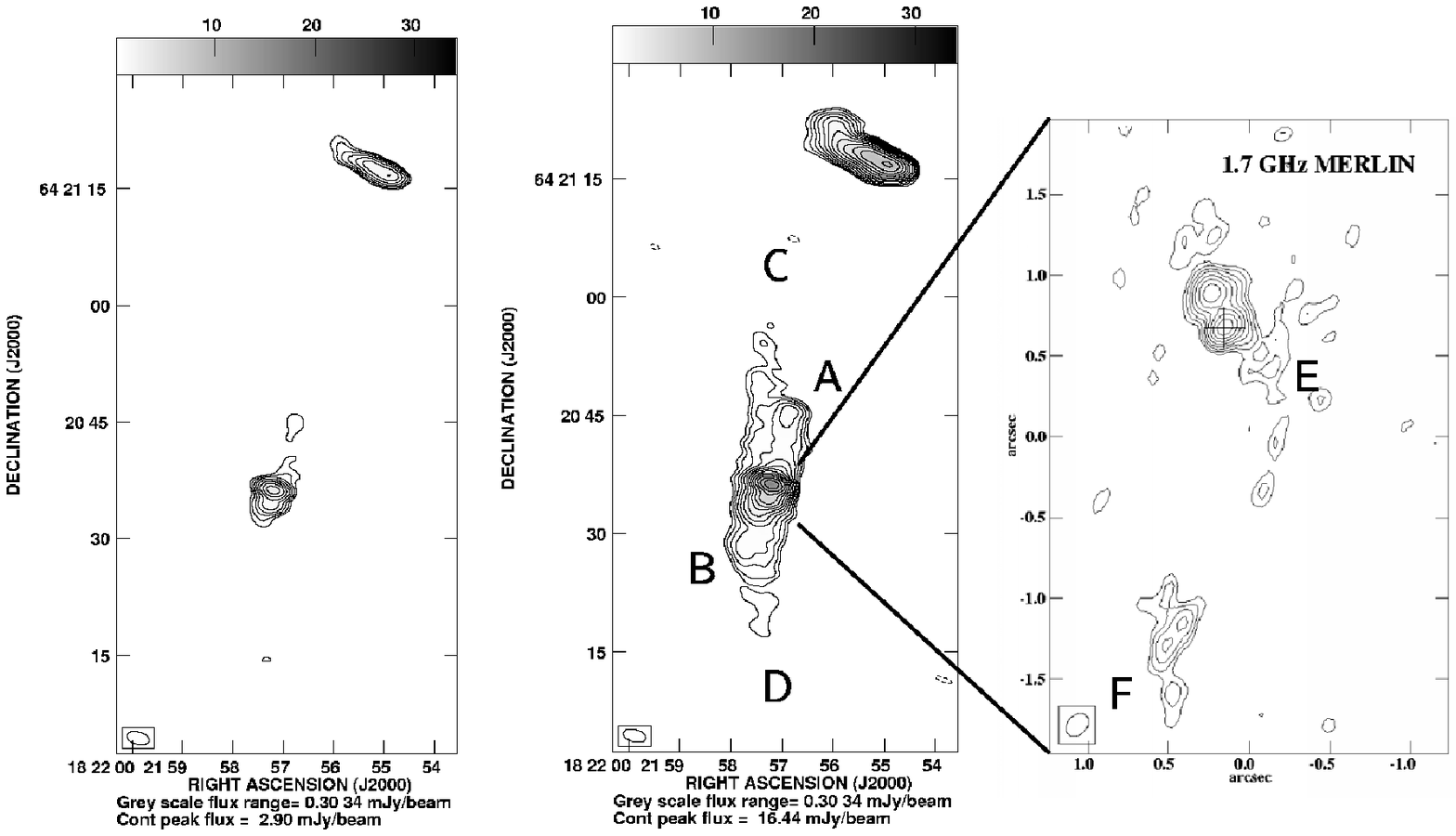}
\caption{The central panel is a VLA image of E\,1821+643 at 1.4\,GHz
with arcsec resolution which shows the quasar to also have an FR\,I
structure extending over 280\,kpc.  The northern source is a known
FR\,I radio galaxy in the same cluster \citep{Lac92}.  The bright
component in this image is shown at higher resolution in the right
image from MERLIN \citep[reproduced from][]{Blu96} to contain
oppositely directed twin jets with evidence for some curvature.  The
{\em left} image shows the effect of redshifting this (most radio
luminous of radio-quiet quasars) out to redshift 1 assuming a spectral
index of 0.9 for the extended emission and making the assumptions
about surface brightness discussed in the text, and observing it at
1.4\,GHz for the same length of time and with the same UV-coverage as
in the central panel.  This is approximately the same as a 5-min
snapshot of the source at its true redshift of 0.297, though the
reality would not be as good because of deconvolution difficulties
arising from significantly poorer UV-plane coverage.
\label{fig:images}
}
\end{figure}

The VLA in its BnA configuration was used to observe E\,1821+643 at
18\,cm for 133 minutes on 14 September 1995; the primary flux
calibrator was 3C\,48 and the phase calibrator was 1842+681.  The data
were reduced using standard procedures using AIPS.
Fig.\,\ref{fig:images} shows that at 18\,cm low-surface brightness
emission (which contributes to a total of 47.7\,mJy associated with
the quasar) is extended over 45$^{\prime\prime}$ (280\,kpc), way
beyond the confines of its host galaxy \citep{Hut91}.  Previously,
extended radio emission had only been known to lie approximately
co-spatially with its host galaxy \citep{Pap95}.  Despite the somewhat
elongated nature of the radio emission \citeauthor{Pap95} found, they
thought it likely that this emission was associated with star
formation within the galaxy; this was concluded because of the
proximity of this quasar to the seemingly tight line of correlation
between the far-infra-red and radio luminosities of Seyfert galaxies
and radio-quiet quasars \citep[e.g.][]{Hel85,Sop91a,Sop91b}.  The
increased radio emission we have discovered lifts this object further
away from that tight correlation.

\section{Results}
\label{sec:results}

The total luminosity of this source at 18\,cm is $10^{24.6}$\,${\rm
W\,Hz^{-1}sr^{-1}}$.  At the resolution of the 18\,cm image, the core
component accounts for 15.3\,mJy/beam, i.e.\ half of the luminosity in
the extended emission.  Its linear extent is 280\,kpc in the assumed
cosmology, and its width is 60\,kpc.  Thus in terms of radio size,
luminosity, the ratio of core to total luminosity, host galaxy and
cluster environment \citep{Lac92}, E\,1821+643 is like other FR\,Is.

\centerline{
\psfig{figure=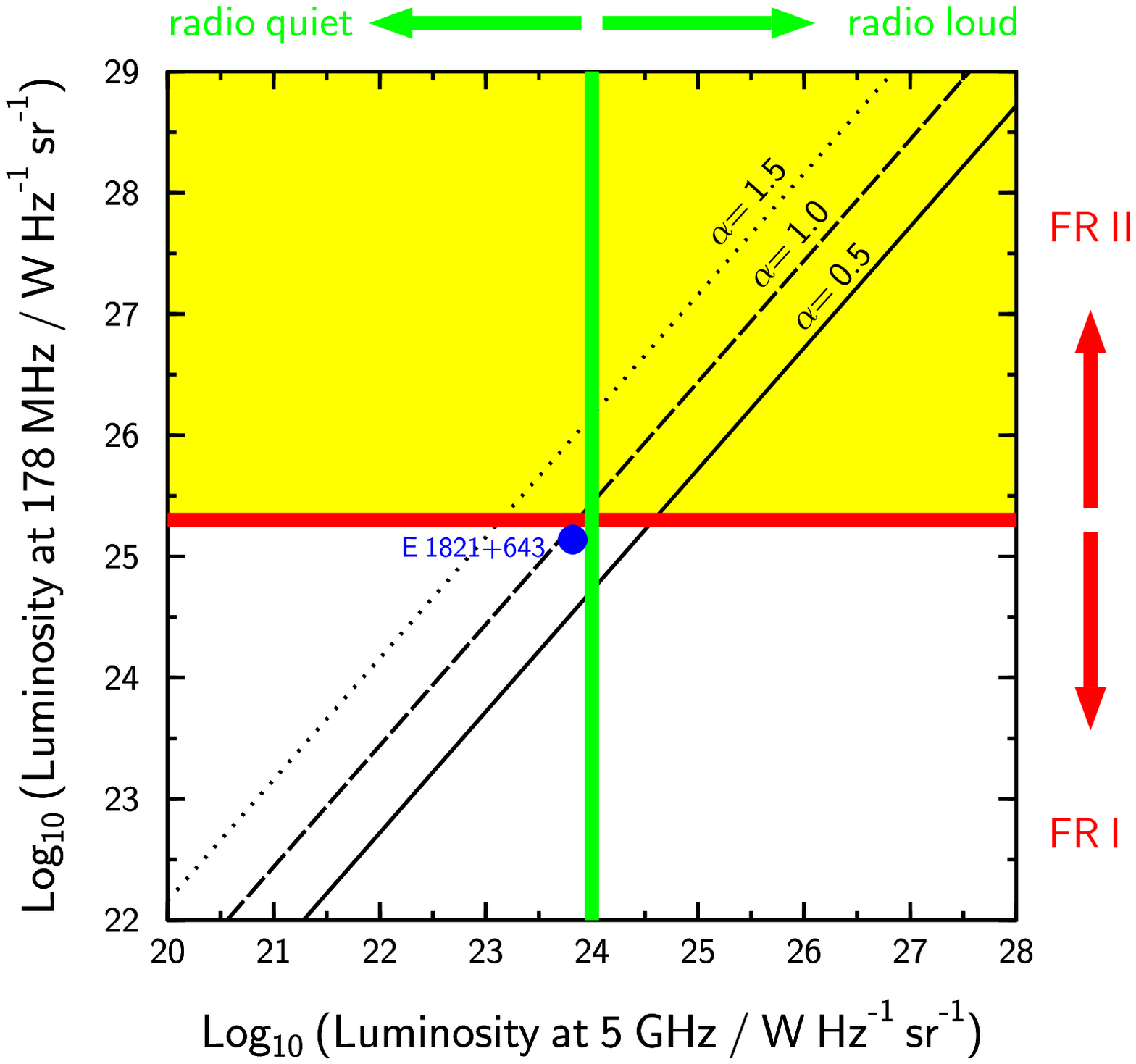,width=9cm,angle=0}
}
\figcaption{Illustration of the relationship between the low-frequency
luminosity ranges associated with the two \citet{Fan74} classes and
the high-frequency luminosity ranges which have been deemed to contain
`radio-loud' or `radio-quiet' quasars by e.g.\ \citet{Kel94}; the
solid, dashed and dotted lines indicate the effective spectral indices
between 178\,MHz and 5\,GHz which must characterize a source appearing
in different parts of this plane.  Even the {\em deepest}
low-frequency surveys to date \citep[e.g.][]{Ril99} with a flux-limit
of 100\,mJy will only find sources at $z > 1$ whose luminosities at
151\,MHz are in the region coloured yellow, i.e.\ in the traditional
FR\,II regime.
\label{fig:rqq_fr}}

It seems likely that the jet observed on sub-arcsec scales by
\citet{Blu96} produces the extended emission.  Fig.\,\ref{fig:images}
shows their MERLIN image in which the jet appears to curve from the
line joining the core to E, round towards F.  This E--F direction is
parallel to A--B on larger scales, which in turn probably represents
more recent emission than that along the direction delineated by C--D.
The axis of the inner MERLIN jet indicates the most recent
directionality of the jet-axis.  Thus, two clues from these images ---
taken together --- may indicate that the jet is precessing: i) the
curvature of the sub-arcsec jet and ii) the similarity in direction of
the jet-axis A--B on scales of 10s of arcsec compared with E--F on
sub-arcsec scales.  This would be in exact analogy to the way the
precessing jets of SS\,433 \citep{Hje81} give rise to a
projected-helical trace on the sky.  It is possible that, as in
SS\,433, the precession of the jet-axis in this radio-quiet quasar
arises from the central engine being a binary.  The changes in
jet-axis direction may be the manifestation of the active nucleus
containing two super-massive black holes in orbit round one another
\citep{Beg80}.

The outermost ends of the extended emission do not appear to end in
shocks (as might be evinced by an abrupt, rather than a gradual,
decrease in surface brightness).  This might suggest that the extended
emission has grown no faster than the ambient sound speed of the IGM.
The sound speed in an ionized gas at temperature $T$ K is given by $
\approx 5 \times 10^{-7} \times c \sqrt{T}$ where $c$ is the speed of
light; for $T = 10^7$K this gives an assumed transonic advance speed
of $0.0015c$.  If this were the constant speed at which the plumes had
emerged from the radio-quiet quasar, then this would have commenced
$\sim 3 \times 10^8$\,yrs ago so the direction of the jet-axis appears
to have remained within a 7\,degree cone for at least $\sim 3 \times
10^8$\,yrs.  With the A--B distance being somewhat under half the C--D
distance, this still gives that the characteristic time for the
stability of the jet axes is $\sim 10^8$\,yrs.  The precession period
of SS\,433 is 163\,days, $\sim 10^8$ times shorter than the likely
period of E\,1821+643; thus the ratio of precession periods in these
very different objects would be similar to the ratio of their masses.

We now consider an estimate of the stability timescale of the optical
emission: \cite{Fri98} found extended line emission
2.5$^{\prime\prime}$ (7.5\,kpc) from the nucleus in E\,1821+643, with
line ratios typical of nuclear photoionization, suggesting stability
over $\sim 10^4$\,yrs.  The width of this emission perpendicular to
the mean jet-axis is similar to the width of the radio emission,
suggesting that the radio source may be interacting with this gas.

\section{The consequences of redshift}
\label{sec:redshift}

We now consider whether extended radio emission is likely to be a
general characteristic of radio-quiet quasars or whether E\,1821+643
is simply an exceptional case.  Radio maps of other radio-quiet
quasars from e.g.\ \citet{Kuk98} and \citet{Whi00} had an on-source
time of only a few minutes, compared with 2\,hrs for the image we
present here: longer integration time brings with it not just a simple
increase in signal-to-noise ratio but also substantially better {\em
UV}-plane coverage.

Have previous observations of radio-quiet quasars simply had
insufficient surface-brightness sensitivity for diffuse extended radio
structures to be detected at even moderate redshifts?  We performed an
experiment similar to one performed by \citet{Nee95} in a different
context: we took our image of E\,1821+643 and calculated what it would
look like if instead of being located at the redshift of 0.297
\citep{Sch92}, it were located at the maximum redshift at which an
otherwise identical but unreddened E\,1821+643 could be detected by
the BQS survey, namely $z \sim 1$.  For constant ${\rm
W\,Hz^{-1}\,kpc^{-2}}$ over physical sizes exceeding the pixel size
(in kpc at any considered $z$), the surface brightness will vary with
redshift as $(1 + z)^{3 + \alpha}$ (where $\alpha$ is the spectral
index), thus the surface brightness of the object at $z = 1$ would be
lower than if it were at $z = 0.297$ by a factor of 5.4.  The
estimated source structure is shown in the left panel of
Fig.\,\ref{fig:images}: at the risk of stating the obvious {\em there
would be no reason to deduce the presence of 100\,kpc-scale emission
on the basis of such an image}.  For sources at moderate to high
redshift, it is difficult to observe any low-luminosity extended
emission.

\section{Jet power and luminosity in FR\,Is and RQQs}
\label{sec:sourcephysics}

The FR\,I radio galaxies in 3C may differ from E\,1821+643 in that the
presence or absence of examples clearly indicating precession requires
MERLIN-scale imaging.  The FR\,I radio galaxies in 3C definitely
differ from E\,1821+643 in that their optical luminosities are $10^3
\times$ lower, though their radio luminosities are similar.

{\em Do any other `radio-quiet' quasars have 100-kpc-scale jets}
(manifested as FR\,Is)?  Figure\,\ref{fig:rqq_fr} clearly demonstrates
that with plausible radio spectral indices, and given current
sensitivity at low-frequency, the relevant part of parameter space is
not sampled, so for the time being this question remains unanswered,
though it is possible that some do have such jets (e.g.\ those with
elliptical hosts) and some do not (e.g.\ those with spiral hosts).
Only if the answer to this question turns out to be negative (after
deeper exposures at low-$z$ with existing facilities, and ultimately
with e-MERLIN, e-VLA, LoFar and SKA, have sampled this parameter
space) would it be possible to posit a bimodality, rather than a
continuity, in the luminosities of the jet output of quasars.

Implicit in the work of \citet{Raw91}, \citet{Mil93} and \citet{Wil99}
was the view that there is a physical dichotomy between `radio-loud'
objects (which channel power into jets of the same order as is
released by accretion) and `radio-quiet' objects (whose jet powers are
a negligible fraction of the bolometric accretion luminosity).  This
view does not rest easily with the optically-powerful E\,1821+643
since it has nearly all of the attributes (\S\ref{sec:results}) of a
low-jet-power FR\,I radio source and is therefore intermediate between
high-power FR\,IIs and classically `radio-quiet' objects.  One way of
retaining a physical dichotomy would be to postulate a different
mapping between jet power and observed low-frequency radio luminosity
in (at least) this FR\,I.  Could FR\,I-like structures from
optically-powerful nuclei emerge from different physical processes?
We speculate that the radio luminosity of some radio-quiet quasars may
be reduced by significant precession in their jet-axes.  This
precession could arise if insufficient time has elapsed for orbiting
black holes in the central engine to have coalesced.

The correlation with radio luminosity of the structural
classifications of \citet{Fan74} is one of the most persistent and
robust correlations in astronomy: there are {\em no} examples of
highly radio-luminous FR\,Is.  This suggests that high jet power is
necessary for a highly collimated non-dissipative jet which is
ultimately capable of forming compact hotspots and a characteristic
FR\,II structure.  As lower jet powers are considered, a jet may be
more likely to disrupt and dissipate within a few kpc of the cores,
characteristic of FR\,Is [see e.g.\ \citet{DeY93}], but giving a low
luminosity source.  The density and inhomogeneity of the environment
into which the jet is expanding determines the exact threshold value
of jet power above which a jet can give rise to an FR\,II and below
which a jet will give rise to an FR\,I.  For a given jet power, if the
jet axis is precessing, this may favour the disruption and hence
dissipation of the jet, lowering the likelihood of an FR\,II.

\section{Concluding remarks}
\label{sec:conc}

Our discovery of an FR\,I radio structure associated with an
unusually-nearby optically-powerful quasar, together with
consideration of the inadequecy of existing observations of
optically-similar objects to reveal similar radio structures, overturn
previous assertions that quasar nuclei cannot be associated with FR\,I
radio structures.  

The {\em large-scale} radio emission of E\,1821+643, elongated way
beyond the quasar's host galaxy, is powered by oppositely directed
jets which appear to precess on timescales which are scaled up in
approximate proportion with the mass of the central engine compared to
the Galactic radio jetted source, SS\,433.  We speculate that the
central engine in this quasar may be composed of two black holes which
have yet to coalesce \citep[cf.][]{Beg80}.

Any precession in the jet-axis of a radio source may smear out its
extended radio emission and promote dissipation in the jet, affecting
its potential to form an FR\,II structure, hindering its detectability
at radio wavelengths.

The ubiquity of precession as a means by which lower radio luminosity
is manifested from highly-accreting yet `radio-quiet' quasars remains
to be established.  This is essential to developing our understanding
of the quasar phenomenon, and beckons the emerging generation of radio
telescopes such as e-MERLIN, e-VLA, LoFar and SKA.

\acknowledgments

K.M.B.\ thanks the Royal Society for a University Research Fellowship.
The VLA is a facility of the NRAO operated by Associated Universities,
Inc., under co-operative agreement with the NSF.  MERLIN is a UK
national facility operated by the University of Manchester on behalf
of PPARC.  We thank the referee for helpful comments.

\vspace{-0.45cm}

\end{document}